\documentclass{article}
\usepackage{amssymb}
\usepackage{amsmath}

\input{tcilatex}
\begin{document}

\begin{center}

{\LARGE Properties of the zeros of the polynomials belonging to the \textit{q%
}-Askey scheme}

\bigskip

$^{\ast }$\textbf{Oksana Bihun}$^{1}$ and $^{+\lozenge }$\textbf{Francesco
Calogero}$^{2}\bigskip $

$^{\ast }$Department of Mathematics, Concordia College\\
 901 8th Str. S,  Moorhead, MN 56562, USA, +1-218-299-4396
\smallskip

$^{+}$Physics Department, University of Rome \textquotedblleft La Sapienza"\\
 p. Aldo Moro, I-00185 ROMA, Italy, +39-06-4991-4372
\smallskip

$^{\lozenge }$Istituto Nazionale di Fisica Nucleare, Sezione di Roma
\smallskip

$^{1}$Corresponding author, obihun@cord.edu

$^{2}$francesco.calogero@roma1.infn.it, francesco.calogero@uniroma1.it

\bigskip

\textit{Abstract}
\end{center}

In this paper we provide properties---which are, to the best of our
knowledge, new---of the zeros of the polynomials belonging to the \textit{q}%
-Askey scheme. These findings include Diophantine relations satisfied by
these zeros when the parameters characterizing these polynomials are
appropriately restricted.
\smallskip


\smallskip

MSC 33C45,
11D41, 
15A18 
\bigskip

\bigskip

\section{Introduction}

In a previous paper \cite{BC2014a} we identified certain new properties of
the $N$ zeros of the polynomials of order $N$ belonging to the Askey scheme.
The main one of these properties identifies an $N\times N$
matrix---explicitly defined in terms of these $N$ zeros and of the
parameters characterizing the polynomial under consideration---which
features $N$ eigenvalues given by neat, explicit formulas, having moreover a
Diophantine connotation when the parameters of the polynomial are suitably
restricted. In this paper we identify somewhat analogous properties of the $%
N $ zeros of the polynomials of order $N$ belonging to the \textit{q}-Askey
scheme. Above and hereafter $N$ is an \textit{arbitrary positive integer},
and $q$ an arbitrary number (possibly even \textit{complex}; of course the
results reported in this paper reproduce--via appropriate developments---the
results of \cite{BC2014a} in the $q\rightarrow 1$ limit).

Let us recall \cite{BC2014a} that \textquotedblleft the properties of the
zeros of polynomials are a core problem of mathematics to which, over time,
an immense number of investigations have been devoted. Nevertheless new
findings in this area continue to emerge, see, for instance, \cite%
{BCD1, BCD2, BCD3, BCD4,BCD5, CI2013,IR2013,BC2013,BC2014,CY1, CY2,BCY,C1,C2,C3},'' and see also the very recent
paper \cite{AT2014}.

The technique used to obtain the results reported below is somewhat
analogous to that employed in our previous paper \cite{BC2014a}, but there
are significant differences, due to the fact that the main tool of our
treatment are now Differential \textit{q}-Difference Equations (D\textit{q}%
DEs) instead of Differential Difference Equations (DDEs). Hence our
treatment below is patterned after that of \cite{BC2014a}; yet a previous
reading of \cite{BC2014a} is by no means mandatory to understand what
follows.

The main findings of this paper are reported in the following Section 2.
They detail properties of the zeros of the Askey-Wilson and \textit{q}-Racah
polynomials, which are the two ``highest'' classes of polynomials belonging
to the \textit{q}-Askey scheme \cite{KS}---so that these polynomials feature 
$5$ arbitrary parameters (including $q\neq 1$), in addition to their degree $%
N$. Analogous properties can of course be obtained, from the results
reported below, for the zeros of the (variously ``named'' \cite{KS})
polynomials belonging to ``lower'' classes of the \textit{q}-Askey scheme,
via the reductions---corresponding to special assignments of the
parameters---that characterize the \textit{q}-Askey scheme \cite{KS}; we
leave this task to the interested reader.

Our findings are proven in Section 3. The definitions and some standard
properties of the Askey-Wilson and \textit{q}-Racah polynomials are reported
in the Appendix, for the convenience of the reader and also to specify our
notation; the reader is advised to glance through this Appendix before
reading the next section, and then to return to it whenever appropriate.

The results that follow are only a consequence of the \textit{explicit} 
\textit{definitions} of the Askey-Wilson and \textit{q}-Racah polynomials
and of the \textit{q}-\textit{difference equations} they satisfy (see the
Appendix); the orthogonality properties that these polynomials also satisfy
play no role, so that the results reported below do not require the
restrictions on the parameters and arguments of these polynomials that are
instead mandatory for the validity of these orthogonality properties and of
other related properties \cite{KS}.

\section{Results}

To formulate our results we refer to the definitions and standard properties
of the Askey-Wilson and \textit{q}-Racah polynomials as reported in the
Appendix, to which the reader should also refer for the notation employed
hereafter. As indicated below, some of these results are immediate
consequences of known properties of the Askey-Wilson and \textit{q}-Racah
polynomials; our main results are proven in the following Section 3.

\subsection{Results for the zeros of the Askey-Wilson polynomials}

\textbf{Notation 2.1}. In this Section 2.1---as in the Askey-Wilson parts of
the Appendix and of the next Section 3---we often use the change of
variables 
\begin{subequations}
\begin{equation}
z=x+\sqrt{x^{2}-1}~,~~~x=\frac{z^{2}+1}{2~z}~,  \label{xz}
\end{equation}%
with $x$ being the argument of the Askey-Wilson polynomial $p_{N}\left(
x\right) \equiv p_{N}\left( a,b,c,d;q;x\right) $ and $z$ being the argument
of the corresponding rational function $P_{N}\left( z\right) \equiv
P_{N}\left( a,b,c,d;q;z\right) $, see (\ref{PpAW}). These relations
correspond to the assignment $x=\cos \theta $, $z=\exp \left( 
\mathbf{i}\theta \right) $ (here and hereafter $\mathbf{i}$ denotes the
imaginary unit, $\mathbf{i}^{2}=-1)$; but of course in this paper the
argument $x$ \ of the Askey-Wilson polynomial is a \textit{complex variable}%
, not restricted to the interval $[-1,1]$ of the real line. And in
all our final formulas only even powers of the square-root $\sqrt{x^{2}-1}$
appear, so the determination of the square-root in the first formula (\ref%
{xz})---and in analogous formulas, see below---is not an issue. Likewise,
the $N$ zeros of the Askey-Wilson polynomial $p_{N}\left( x\right) $ are
denoted as $\bar{x}_{n},~p_{N}\left( \bar{x}_{n}\right) =0$, and the $N$
quantities%
\begin{equation}
\bar{z}_{n}=\bar{x}_{n}+\sqrt{\bar{x}_{n}^{2}-1}~,~~~\bar{x}_{n}=\frac{\bar{z%
}_{n}^{2}+1}{2~\bar{z}_{n}}~,  \label{xznbar}
\end{equation}%
are $N$ zeros of the rational function $P_{N}\left( z\right) $, $P_{N}\left( 
\bar{z}_{n}\right) =0$. $\square $

\textbf{Proposition 2.1}. Let $\bar{x}_{n}\equiv \bar{x}_{n}\left(
a,b,c,d;q;N\right) $ (with $n=1,2,...,N)$ be the $N$ zeros of the
Askey-Wilson polynomial $p_{N}\left( x\right) \equiv p_{N}\left(
a,b,c,d;q;x\right) $ of degree $N$ in $x$ (with $N$ an arbitrary positive
integer), so that (up to an irrelevant multiplicative constant $C_{N}$%
\textbf{) }$p_{N}\left( x\right) =C_{N}~\prod\nolimits_{n=1}^{N}\left( x-%
\bar{x}_{n}\right) $ ; and let $\bar{z}_{n}$ be related to $\bar{x}_{n}$by (%
\ref{xznbar}).

Then there hold the $N$\ algebraic equations 
\end{subequations}
\begin{subequations}
\begin{equation}
A\left( \bar{z}_{n}\right) ~p_{N}\left( \frac{q^{2}~\bar{z}_{n}^{2}+1}{2~q~%
\bar{z}_{n}}\right) +A\left( \frac{1}{\bar{z}_{n}}\right) ~p_{N}\left( \frac{%
\bar{z}_{n}^{2}+q^{2}}{2~q~\bar{z}_{n}}\right) =0~,n=1,\ldots ,N~,
\label{EqA21a}
\end{equation}%
or, equivalently,%
\begin{equation}
A\left( \bar{z}_{n}\right) ~\prod\limits_{m=1}^{N}\left( q\bar{z}_{n}+\frac{1%
}{q\bar{z}_{n}}-\bar{z}_{m}-\frac{1}{\bar{z}_{m}}\right) +\left[ \left( \bar{%
z}_{s}\rightarrow \frac{1}{\bar{z}_{s}}\right) \right] =0~,~n=1,\ldots ,N~,
\label{EqA21b}
\end{equation}%
where%
\begin{equation}
A\left( z\right) \equiv A\left( a,b,c,d;q;z\right) =\frac{\left( 1-az\right)
~\left( 1-bz\right) ~\left( 1-cz\right) ~\left( 1-dz\right) }{\left(
1-z^{2}\right) ~\left( 1-qz^{2}\right) }  \label{AA}
\end{equation}%
and the symbol $+\left[ \left( \bar{z}_{s}\rightarrow \frac{1}{\bar{z}_{s}}%
\right) \right] $ denotes the addition of everything that comes before it,
with $\bar{z}_{s}$ replaced by $\frac{1}{\bar{z}_{s}}$ for all $s=1,\ldots
,N $. $\square $

\textbf{Remark 2.1}\textit{.} It is evident that after the substitution $%
\bar{x}_{n}=\cos \bar{\theta}_{n}$ and $\bar{z}_{n}=\exp \left( \mathbf{i}%
\bar{\theta}_{n}\right) $, see~(\ref{xznbar}), the left-hand side of (\ref%
{EqA21b}) becomes an \textit{even} function of each $\bar{\theta}_{s}$, $%
s=1,\ldots ,N$, hence a function of $\bar{x}_{1},\ldots ,\bar{x}_{N}$. This
is why after the substitution $\bar{z}_{n}=\bar{x}_{n}+\sqrt{\bar{x}%
_{n}^{2}-1}$ in the left-hand sides of~(\ref{EqA21a}) and~(\ref{EqA21b}) all
the square roots disappear, regardless of the determination of each square
root~$\sqrt{\bar{x}_{n}^{2}-1}$, as long as it is consistent, for each $n$,
throughout the treatment. $\square $

This result is proved in the Appendix, see (\ref{Prop21}).

The following proposition is our main result for the $N$ zeros $\bar{x}_{n}$
of the Askey-Wilson polynomials.

\textbf{Proposition 2.2}. Let $\bar{x}_{n}\equiv \bar{x}_{n}\left(
a,b,c,d;q;N\right) $ (with $n=1,2,...,N)$ be the $N$ zeros of the
Askey-Wilson polynomial $p_{N}\left( x\right) \equiv p_{N}\left(
a,b,c,d;q;x\right) $ of degree $N$ in $x$ (with $N$ an arbitrary positive
integer), so that (up to an irrelevant multiplicative constant $C_{N}$) $%
p_{N}\left( x\right) =C_{N}~\prod\nolimits_{n=1}^{N}\left( x-\bar{x}%
_{n}\right) $ ; and let $\bar{z}_{n}$ be related to $\bar{x}_{n}$ by (\ref%
{xznbar}). Define the $N\times N$ matrix $\underline{M}\equiv \underline{M}%
\left( a,b,c,d;q;N;\underline{\bar{x}}\right) $, componentwise, as follows: 
\end{subequations}
\begin{subequations}
\label{M}
\begin{eqnarray}
&&M_{nn}\equiv M_{nn}\left( a,b,c,d,;q;N;\underline{\bar{x}}\right)   \notag
\\ \notag
&=&\frac{(q-1)}{2q^{N}}\Bigg\{\Bigg[\frac{2\bar{z}_{n}^{2}}{\bar{z}_{n}^{2}-1%
}G(\bar{z}_{n})\sum_{m=1,m\neq n}^{N}\Big(-\frac{q}{\bar{z}_{m}-q\bar{z}_{n}}%
+\frac{q\bar{z}_{m}}{q\bar{z}_{n}\bar{z}_{m}-1} \\
&&+\frac{1}{\bar{z}_{m}-\bar{z}_{n}}-\frac{\bar{z}_{m}}{\bar{z}_{n}\bar{z}%
_{m}-1}\Big)+\frac{2\bar{z}_{n}^{2}G^{\prime }(\bar{z}_{n})}{\bar{z}%
_{n}^{2}-1}\Bigg]\prod_{\ell =1,\ell \neq n}^{N}K(\bar{z}_{n},\bar{z}_{\ell
})  \notag \\
&&+\left[ \left( \bar{z}_{s}\rightarrow \frac{1}{\bar{z}_{s}}\right) \right] %
\Bigg\}
\end{eqnarray}%
for $n=1,\ldots ,N,$ and 
\begin{eqnarray}
&&M_{nm}\equiv M_{nm}\left( a,b,c,d,;q;N;\underline{\bar{x}}\right)   \notag
\\ \notag
&=&\frac{(q-1)}{2q^{N}}\Bigg\{\frac{2\bar{z}_{m}^{2}}{\bar{z}_{m}^{2}-1}G(%
\bar{z}_{n})\Big[\frac{1}{\bar{z}_{m}-q\bar{z}_{n}}+\frac{q\bar{z}_{n}}{q%
\bar{z}_{n}\bar{z}_{m}-1} \\
&&-\frac{1}{\bar{z}_{m}-\bar{z}_{n}}-\frac{\bar{z}_{n}}{\bar{z}_{n}\bar{z}%
_{m}-1}\Big]\prod_{\ell =1,\ell \neq n}^{N}K(\bar{z}_{n},\bar{z}_{\ell })+%
\left[ \left( \bar{z}_{s}\rightarrow \frac{1}{\bar{z}_{s}}\right) \right] %
\Bigg\}
\end{eqnarray}%
for $n,m=1,...,N~\ $with $n\neq m$. In these formulas%
\begin{equation}
G(\bar{z}_{n})=A(\bar{z}_{n})~\left( q\bar{z}_{n}-\frac{1}{\bar{z}_{n}}%
\right) ~,  \label{eq:GG}
\end{equation}%
\begin{equation}
G^{\prime }(z)=\frac{d}{dz}G(z)~,
\end{equation}%
$A\equiv A\left( a,b,c,d;q;z\right) $ is defined by~(\ref{AA}), 
\begin{equation}
K(\bar{z}_{n},\bar{z}_{m})=\frac{(\bar{z}_{m}-q\bar{z}_{n})~(q\bar{z}_{n}%
\bar{z}_{m}-1)}{(\bar{z}_{m}-\bar{z}_{n})~(\bar{z}_{n}\bar{z}_{m}-1)}
\label{eq:KK}
\end{equation}%
for $n,m=1,\ldots ,N$ with $n\neq m$, and the symbol $+\left[ \left( \bar{z}%
_{s}\rightarrow \frac{1}{\bar{z}_{s}}\right) \right] $ indicates addition of
everything that comes before it, within the curly brackets, with $\bar{z}_{s}
$ replaced by $\frac{1}{\bar{z}_{s}}$ for all $s=1,\ldots ,N$.

Then this $N\times N$ matrix $\underline{M}$ has the $N$ eigenvalues%
\begin{eqnarray}
\mu _{n} &\equiv &\mu _{n}\left( abcd;q;N\right) =q^{-N}\left(
1-q^{n}\right) \left( 1-abcd~q^{2N-1-n}\right)  \notag \\
&=&\left( q^{-N}+abcd~q^{N-1}-q^{n-N}-abcd~q^{N-1-n}\right) ~,  \notag \\
n &=&1,2,...,N~.~\square
\end{eqnarray}

It is evident that the components of the matrix $M$ in \textbf{Proposition
2.2} are functions of $\bar{x}_{1},\ldots ,\bar{x}_{N}$, see \textbf{Remark
2.1}.

Some immediate corollaries of \textbf{Proposition 2.2} are worth a mention.

\textbf{Corollary 2.2.1}. If $abcd$ and $q$ are both \textit{rational}
numbers, the $N$ eigenvalues of the $N\times N$ matrix $\underline{M}$ (see (%
\ref{M})) are all \textit{rational} numbers.~$\square $.

This is a remarkable Diophantine property.

\textbf{Corollary 2.2.2}. The $N\times N$ matrix $\underline{M}$---which
depends of course on the $4$ \textit{a priori} arbitrary parameters $a,b,c,d$%
, explicitly via $A\left( z\right) ,$ see (\ref{AA}), and implicitly via the
dependence on these $4$ parameters of the $N$ zeros $\bar{x}_{n}\equiv \bar{x%
}_{n}\left( a,b,c,d;q;N\right) $ of the Askey-Wilson polynomials $%
p_{N}\left( a,b,c,d;q;x\right) $, see (\ref{M}), is \textit{isospectral}
under any variation of these $4$ parameters which does not change the value
of their product $abcd$.~$\square $

\textbf{Corollary 2.2.3}. Several identities satisfied by the $N$ zeros $%
\bar{x}_{n}\equiv \bar{x}_{n}\left( a,b,c,d;q;N\right) $ of the Askey-Wilson
polynomial $p_{N}\left( x\right) \equiv p_{N}\left( a,b,c,d;q;x\right) $ are
implied by the following standard consequences of \textbf{Proposition 2.2}: 
\end{subequations}
\begin{subequations}
\begin{eqnarray}
\text{trace}\left[ \left( \underline{M}\right) ^{k}\right]
&=&\sum_{n=1}^{N}\left( q^{-N}+abcd~q^{N-1}-q^{n-N}-abcd~q^{N-1-n}\right)
^{k}~,  \notag \\
k &=&1,2,3,...~,
\end{eqnarray}

\begin{equation}
\det \left[ \underline{M}\right] =q^{-N^{2}}~\left( q;q\right) _{N}~\left(
abcd~q^{N-1};q\right) _{N}~.
\end{equation}%
In particular, for $k=1,$ the first of these two formulas reads as follows: 
\begin{equation}
\sum_{n=1}^{N}\left( M_{nn}\right) =N\left( q^{-N}+abcd~q^{N-1}\right)
+\left( \frac{1-q^{-N}}{1-q}\right) \left( q+abcd~q^{N-1}\right) ~.~\square
\end{equation}
\end{subequations}

\subsection{Results for the zeros of the \textit{q}-Racah polynomials}

Let $\bar{z}_{n}\equiv \bar{z}_{n}\left( \alpha ,\beta ,\gamma ,\delta
;q;N\right) $, where $n=1,2,...,N$, be the $N$ zeros of the \textit{q}-Racah
polynomial $R_{N}\left( x\right) \equiv R_{N}\left( \alpha ,\beta ,\gamma
,\delta ;q;z\right) $ of degree $N$ in $z$ (with $N$ an arbitrary positive
integer), where the variables $x$ and $z$ are related by formula~(\ref%
{qRacah}). Thus, up to an irrelevant multiplicative constant $C_{N}$, $%
R_{N}\left( z\right) =C_{N}~\prod\nolimits_{n=1}^{N}\left( z-\bar{z}%
_{n}\right) $, see~(\ref{FacRacah}). Let the $2N$ numbers $\bar{z}%
_{n}^{\left( \pm \right) }$ be defined as follows:%
\begin{equation}
\bar{z}_{n}^{\left( \pm \right) }=q^{\pm 1}\bar{z}_{n}\pm \left( \frac{%
1-q^{2}}{2q}\right) \left( \bar{z}_{n}-\sqrt{\bar{z}_{n}^{2}-4\gamma \delta q%
}\right) ,\;\;n=1,\ldots ,N~,  \label{zn+-}
\end{equation}%
see (\ref{z+-}) (of course with the same determination of the square root as
in (\ref{Z})).

A result concerning the $N$ zeros of the \textit{q}-Racah polynomials of
degree $N$ reads then as follows.

\textbf{Proposition 2.3}. The $N$ zeros $\bar{z}_{n}\equiv \bar{z}_{n}\left(
\alpha ,\beta ,\gamma ,\delta ;q;N\right) $, where $n=1,\ldots ,N$, of the $N
$-th degree $q$-Racah polynomial $R_{N}\left( \alpha ,\beta ,\gamma ,\delta
;q;z\right) $ satisfy the following $N$ algebraic equations: 
\begin{subequations}
\label{Prop23}
\begin{equation}
B\left( \bar{z}_{n}\right) ~R_{N}(\bar{z}_{n}^{\left( +\right) })+D\left( 
\bar{z}_{n}\right) ~R_{N}(\bar{z}_{n}^{\left( -\right) })=0~,~~~n=1,...,N~,
\label{Prop23a}
\end{equation}%
or, equivalently,%
\begin{equation}
B\left( \bar{z}_{n}\right) ~\prod\limits_{m=1}^{N}\left( \bar{z}_{n}^{\left(
+\right) }-\bar{z}_{m}\right) +D\left( \bar{z}_{n}\right)
~\prod\limits_{m=1}^{N}\left( \bar{z}_{n}^{\left( -\right) }-\bar{z}%
_{m}\right) =0~,~~~n=1,...,N~,  \label{Prop23b}
\end{equation}%
were the numbers $\bar{z_{n}}^{\left( \pm \right) }$ are defined by~(\ref%
{zn+-}) and the functions $B\left( z\right) $ respectively $D\left( z\right) 
$ are defined by (\ref{B}) respectively~(\ref{D},~\ref{Z}). Note that this
result holds independently of which determination is taken for the square
root in the above definition of $\bar{z}_{n}^{\left( \pm \right) }$ and in (%
\ref{Z}), provided of course it is the same.~$\square $

This \textbf{Proposition 2.3} is proven in the Appendix, see (\ref{48a}).

The following proposition is our main result for the zeros of \textit{q}%
-Racah polynomials.

\textbf{Proposition 2.4}. Let $\bar{z}_{n}\equiv \bar{z}_{n}\left( \alpha
,\beta ,\gamma ,\delta ;q;N\right) $, where $n=1,2,...,N$, be the $N$ zeros
of the \textit{q}-Racah polynomial $R_{N}\left( \alpha ,\beta ,\gamma
,\delta ;q;z\right) $ of degree $N$ in $z$ (with $N$ an arbitrary positive
integer). Define the $N\times N$ matrix $\underline{L}\equiv \underline{L}%
\left( \alpha ,\beta ,\gamma ,\delta ;q;N;\underline{\bar{z}}\right) $,
componentwise, as follows: 
\end{subequations}
\begin{subequations}
\label{L}
\begin{eqnarray}
&&L_{nn}=\Bigg[B^{\prime }(\bar{z}_{n})(\bar{z}_{n}^{(+)}-\bar{z}_{n}) 
\notag \\
&&+B(\bar{z}_{n})\left( C^{(+)}(\bar{z}_{n})-1+(\bar{z}_{n}^{(+)}-\bar{z}%
_{n})\sum_{m=1,m\neq n}^{N}W^{(+)}(\bar{z}_{n},\bar{z}_{m})\right) \Bigg]%
\prod_{\ell =1,\ell \neq n}^{N}\frac{\bar{z}_{n}^{(+)}-\bar{z}_{\ell }}{\bar{%
z}_{n}-\bar{z}_{\ell }}  \notag \\
&&+\Bigg[D^{\prime }(\bar{z}_{n})(\bar{z}_{n}^{(-)}-\bar{z}_{n})  \notag \\
&&+D(\bar{z}_{n})\left( C^{(-)}(\bar{z}_{n})-1+(\bar{z}_{n}^{(-)}-\bar{z}%
_{n})\sum_{m=1,m\neq n}^{N}W^{(-)}(\bar{z}_{n},\bar{z}_{m})\right) \Bigg]%
\prod_{\ell =1,\ell \neq n}^{N}\frac{\bar{z}_{n}^{(-)}-\bar{z}_{\ell }}{\bar{%
z}_{n}-\bar{z}_{\ell }},  \notag \\
&&n=1,...,N~,  \label{Lnn}
\end{eqnarray}%
\begin{eqnarray}
&&L_{nm}=B(\bar{z}_{n})\left( \frac{\bar{z}_{n}^{(+)}-\bar{z}_{n}}{\bar{z}%
_{n}-\bar{z}_{m}}\right) ^{2}\prod_{\ell =1\,\ell \neq n,m}^{N}\frac{\bar{z}%
_{n}^{(+)}-\bar{z}_{\ell }}{\bar{z}_{n}-\bar{z}_{\ell }}  \notag \\
&&+D(\bar{z}_{n})\left( \frac{\bar{z}_{n}^{(-)}-\bar{z}_{n}}{\bar{z}_{n}-%
\bar{z}_{m}}\right) ^{2}\prod_{\ell =1\,\ell \neq n,m}^{N}\frac{\bar{z}%
_{n}^{(-)}-\bar{z}_{\ell }}{\bar{z}_{n}-\bar{z}_{\ell }}%
,~n,m=1,...,N,~n\neq m,  \label{Lnm}
\end{eqnarray}%
where, as above, the functions $B\left( z\right) $ respectively $D\left(
z\right) $ are defined by (\ref{B}) respectively~(\ref{D},~\ref{Z}), $%
B^{\prime }(z)=\frac{d}{dz}B(z)$, 
\begin{equation*}
C^{(\pm )}(\bar{z}_{n})=\frac{d\bar{z}_{n}^{(+)}}{d\bar{z}_{n}}=q^{\pm 1}\pm 
\frac{1-q^{2}}{2q}\left( 1-\frac{\bar{z}_{n}}{\sqrt{\bar{z}_{n}^{2}-4\gamma
\delta q}}\right) ,
\end{equation*}%
\begin{equation*}
W^{(\pm )}(\bar{z}_{n},\bar{z}_{m})=\frac{C^{(\pm )}(\bar{z}_{n})(\bar{z}%
_{n}-\bar{z}_{m})-\bar{z}_{n}^{(\pm )}+\bar{z}_{m}}{(\bar{z}_{n}-\bar{z}%
_{m})(\bar{z}_{n}^{(\pm )}-\bar{z}_{m})},
\end{equation*}%
and the numbers $\bar{z}_{n}^{\left( \pm \right) }$ are defined by~(\ref%
{zn+-}). Then this $N\times N$ matrix $\underline{L}$ has the $N$ eigenvalues%
\begin{eqnarray}
\lambda _{n} &\equiv &\lambda _{n}\left( \alpha \beta ;q;N\right)
=q^{-N}\left( 1-q^{m}\right) \left( 1-\alpha \beta ~q^{2N-m+1}\right) ~, 
\notag \\
n &=&1,2,...,N~.~\square 
\end{eqnarray}

Some immediate corollaries of \textbf{Proposition 2.4} are worth a mention.

\textbf{Corollary 2.4.1}. If $\alpha \beta $ and $q$ are both \textit{%
rational} numbers, the $N$ eigenvalues of the $N\times N$ matrix $\underline{%
L}$ (see (\ref{L})) are all \textit{rational} numbers.~$\square $.

This is a remarkable Diophantine property.

\textbf{Corollary 2.4.2}. The $N\times N$ matrix $\underline{L}$---which
depends of course on the $4$ \textit{a priori} arbitrary parameters $\alpha
,\beta ,\gamma ,\delta $, explicitly via $B\left( z\right) $ and $D\left(
z\right) $ (see (\ref{B}) and (\ref{D}) with (\ref{Z})), and implicitly via
the dependence on these $4$ parameters of the $N$ zeros $\bar{z}_{n}\equiv 
\bar{z}_{n}\left( \alpha ,\beta ,\gamma ,\delta ;q;N\right) $ of the \textit{%
q}-Racah polynomial $R_{N}\left( \alpha ,\beta ,\gamma ,\delta ;q;z\right) $%
: see (\ref{L})---is \textit{isospectral} under any variation of these $4$
parameters which does not change the value of the product $\alpha \beta $.~$%
\square $

\textbf{Corollary 2.4.3}. Several identities satisfied by the $N$ zeros $%
\bar{z}_{n}\equiv \bar{z}_{n}\left( \alpha ,\beta ,\gamma ,\delta
;q;N\right) $ of the \textit{q}-Racah polynomial $R_{N}\left( \alpha ,\beta
,\gamma ,\delta ;q;z\right) $ are implied by the following standard
consequences of \textbf{Proposition 2.4}: 
\end{subequations}
\begin{subequations}
\begin{equation}
\text{trace}\left[ \left( \underline{L}\right) ^{k}\right]
=q^{-kN}~\sum_{m=1}^{N}\left[ \left( 1-q^{m}\right) \left( 1-\alpha \beta
~q^{2N-m+1}\right) \right] ^{k}~,~~~k=1,2,3,...~,
\end{equation}

\begin{equation}
\det \left[ \underline{L}\right] =q^{-N^{2}}~\left( q;q\right) _{N}~\left(
\alpha \beta q^{N+1};q\right) _{N}~.
\end{equation}%
In particular, for $k=1,$ the first of these two formulas reads as follows: 
\begin{equation}
\sum_{n=1}^{N}\left( L_{nn}\right) =N\left( q^{-N}+\alpha \beta
~q^{N+1}\right) +\frac{q\left( 1-q^{-N}\right) \left( 1+\alpha \beta
~q^{N}\right) }{1-q}~.~\square
\end{equation}
\section{Proof of the main results}

In this Section 3 we prove our main results for the zeros of the
Askey-Wilson and \textit{q}-Racah polynomials.

\subsection{Askey-Wilson polynomials}

Let $\Psi _{N}\left( z;t\right) \equiv \Psi _{N}\left( a,b,c,d;q;z;t\right) $
be a rational function of the (generally \textit{complex}) variable $z$,
which satisfies the Differential-\textit{q-}Difference Equation (D\textit{q}%
DE) 
\end{subequations}
\begin{equation}
\frac{\partial ~\Psi _{N}\left( z;t\right) }{\partial ~t}=\left[ \left(
q^{-N}-1\right) ~\left( 1-abcd~q^{N-1}\right) -Q\right] ~\Psi _{N}\left(
z,t\right) ~,  \label{DqDE}
\end{equation}%
with the ($t$-independent) \textit{q}-differential operator $Q$ acting on
the variable $z$ as defined by (\ref{Q}). And let us assume that this
rational function can be expressed (as confirmed below) by the following
linear superposition with $t$-dependent coefficients $c_{m}\left( t\right) $
of the $(N+1)$ rational functions $P_{N-m}\left( z\right) \equiv P_{N-m}\left(
a,b,c,d;q;z\right) $ defined by (\ref{PNz}), with $m=0,1,2,...,N$: 
\begin{equation}
\Psi _{N}\left( z,t\right) =\sum_{m=0}^{N}\left[ c_{m}\left( t\right)
~P_{N-m}\left( z\right) \right] ~.
\end{equation}

It is then plain---see the eigenvalue equation (\ref{Qeigen}), of course
with $N$ replaced by $(N-m)$---that the $t$-evolution of the $(N+1)$
coefficients $c_{m}\left( t\right) $ is characterized by the following
system of ODEs, 
\begin{eqnarray}
\dot{c}_{m}\left( t\right) &=&\left[ \left( q^{-N}-1\right) ~\left(
1-abcd~q^{N-1}\right) \right.  \notag \\
&&\left. -\left( q^{m-N}-1\right) ~\left( 1-abcd~q^{N-m-1}\right) \right]
~c_{m}\left( t\right)  \notag \\
&=&q^{-N}~\left( 1-q^{m}\right) \left( 1-abcd~q^{2N-1-m}\right) ~c_{m}\left(
t\right) ~,  \notag \\
m &=&0,1,,2...,N~,  \label{Eqcm}
\end{eqnarray}%
where the superimposed dot denotes of course a $t$-differentiation. Hence 
\begin{subequations}
\label{cmt}
\begin{equation}
c_{0}\left( t\right) =c_{0}\left( 0\right) ~,
\end{equation}%
\begin{equation}
c_{m}\left( t\right) =\exp \left( \mu _{m}~t\right) ~c_{m}\left( 0\right)
~,~~~m=1,2,...,N~,
\end{equation}%
where%
\begin{eqnarray}
\mu _{m} &\equiv &\mu _{m}\left( abcd;q;N\right) =q^{-N}\left(
1-q^{m}\right) \left( 1-abcd~q^{2N-1-m}\right) ~,  \notag \\
m &=&1,2,...,N~,
\end{eqnarray}%
implying%
\begin{equation}
\Psi \left( z;t\right) =c_{0}\left( 0\right) ~P_{N}\left( z\right)
+\sum_{m=1}^{N}\left[ \exp \left( \mu _{m}~t\right) ~c_{m}\left( 0\right)
~P_{N-m}\left( z\right) \right] ~.  \label{Psizt}
\end{equation}

It is now clear (see (\ref{qDE})) that the \textquotedblleft
equilibrium''---i.e., $t$-independent---solution $\bar{\Psi}_{N}\left(
z\right) \equiv \bar{\Psi}_{N}\left( a,b,c,d;q;z\right) $ of D\textit{q}DE (%
\ref{DqDE}) reads 
\end{subequations}
\begin{subequations}
\begin{equation}
\bar{\Psi}_{N}\left( z\right) =c_{0}\left( 0\right) ~P_{N}\left( z\right) ~,
\label{PsiEqui}
\end{equation}%
corresponding to the \textquotedblleft equilibrium'' (i. e., $t$-independent)
solution of the dynamical system (\ref{Eqcm}) which clearly reads%
\begin{equation}
\bar{c}_{0}\left( t\right) =\bar{c}_{0}\left( 0\right) ~;~~~\bar{c}%
_{m}\left( t\right) =0~,m=1,2,...,N~.
\end{equation}

It is at this stage convenient to perform---as in the Appendix---the change
of variables (\ref{xz}) from $z$ to $x$ and viceversa, implying 
\end{subequations}
\begin{subequations}
\begin{equation}
\Psi _{N}\left( z;t\right) =\psi _{N}\left( \frac{z^{2}+1}{2~z};t\right)
~,~~~\psi _{N}\left( x;t\right) =\Psi _{N}\left( x+ \sqrt{x^{2}-1};t\right)
~,  \label{Psipsi}
\end{equation}%
so that (\ref{Psizt}) becomes 
\begin{equation}
\psi \left( x;t\right) =c_{0}\left( 0\right) ~p_{N}\left( x\right)
+\sum_{m=1}^{N}\left[ \exp \left( \mu _{m}~t\right) ~c_{m}\left( 0\right)
~p_{N-m}\left( x\right) \right] ~,  \label{psixt}
\end{equation}%
where $p_{m}\left( x\right) $ is now the Askey-Wilson polynomial of degree $%
m $ (see (\ref{AWpolx}) and (\ref{Psipsi})). It is thus seen that $\psi
\left( x;t\right) $ is a ($t$-dependent) polynomial of degree $N$ in the
variable $x $, and the corresponding \textit{equilibrium}  solution of (\ref{DqDE}) is, up to
the arbitrary multiplicative constant $c_{0}\left( 0\right) ,$ just the
Askey-Wilson polynomial of degree $N$, 
\end{subequations}
\begin{equation}
\bar{\psi}_{N}\left( x\right) =\bar{c}_{0}\left( 0\right) ~p_{N}\left(
x\right)
\end{equation}%
(see (\ref{PsiEqui})).

Next, let us introduce the $N$ zeros $x_{n}\left( t\right) \equiv
x_{n}\left( a,b,c,d;q;N;t\right) $ of the polynomial $\psi \left( x;t\right) 
$ of degree $N$ in $x$, by setting 
\begin{equation}
\psi \left( x;t\right) =C_{N}~\prod\limits_{n=1}^{N}\left[ x-x_{n}\left(
t\right) \right] ~.  \label{Defxnt}
\end{equation}%
Here $C_{N}$ is an arbitrary constant that plays no role in the following;
it is of course proportional to $c_{0}\left( 0\right) $, and the computation
(from (\ref{psixt}), (\ref{Defxnt}) and (\ref{AWpolx})) of the
proportionality factor can be left to the very diligent reader.

Let us now investigate the $t$-evolution of the $N$ zeros $x_{n}\left(
t\right) $, as implied by the Differential \textit{q}-Difference Equation
satisfied by $\psi _{N}\left( x;t\right) ,$ which obtains from the
D\textit{q}DE (\ref{DqDE}) satisfied by $\Psi \left( z;t\right) $ via the
change of variables (\ref{Psipsi}):
\begin{equation}
\frac{\partial ~\psi _{N}\left( x;t\right) }{\partial ~t}=\left[ \left(
q^{-N}-1\right) ~\left( 1-abcd~q^{N-1}\right) -Q\right] ~\psi _{N}\left(
x;t\right) ~.  \label{DDEpsi}
\end{equation}%
The action of the operator $Q$ on functions of the variable $x$ is
given by the formula (see (\ref{Q}) and (\ref{Changexz})) 
\begin{subequations}
\label{QQx}
\begin{equation}
Q~f\left( x\right) =\left[ A\left( z\right) ~\delta _{q}^{\left( +\right)
}+A\left( z^{-1}\right) ~\delta _{q}^{\left( -\right) }-A\left( z\right)
-A\left( z^{-1}\right) \right] ~f\left( x\right)   \label{Qx}
\end{equation}%
with%
\begin{equation}
\delta _{q}^{\left( \sigma \right) }~f\left( x\right) =f\left( q^{\sigma
}x+\sigma \frac{1-q^{2}}{2qz}\right) ~,~~~\sigma =\pm 1~,
\end{equation}%
where the variable $z$ in (\ref{Qx}) is related to the variable $x$
via (\ref{xz}). It can be verified that the right-hand side of (\ref{Qx}) is
an even function of $\theta $ (defined by $z=\exp \left( i\theta \right) )$
hence a function of $x$, and so~--- loosely speaking~--- it makes no
difference to set $z=x+\sqrt{x^{2}-1}$ or $z=x-\sqrt{x^{2}-1}$ in this
definition of the two operators $\delta _{q}^{\left( \pm \right) }$,
provided of course the same convention is used throughout (see \textbf{%
Notation 2.1}).

It is plain from (\ref{Defxnt})---by logarithmic $t$-differentiation---that 
\end{subequations}
\begin{eqnarray}
\frac{\partial ~\psi _{N}\left( x;t\right) }{\partial t}&=&-\psi _{N}\left(
x;t\right) ~\sum_{m=1}^{N}\frac{\dot{x}_{m}\left( t\right) }{x-x_{m}\left(
t\right) }\notag\\
&=&-C_{N}~\sum_{m=1}^{N}\left\{ \dot{x}_{m}\left( t\right)
\prod\limits_{\ell =1,~\ell \neq m}^{N}\left[ x-x_{\ell }\left( t\right) %
\right] \right\} ~.
\end{eqnarray}%
Hence%
\begin{equation}
\left. \frac{\partial ~\psi _{N}\left( x;t\right) }{\partial t}\right\vert
_{x=x_{n}\left( t\right) }=-C_{N}~\left\{ \dot{x}_{n}\left( t\right)
~\prod\limits_{\ell =1,~\ell \neq n}^{N}\left[ x_{n}\left( t\right) -x_{\ell
}\left( t\right) \right] \right\} ~,
\end{equation}%
and by setting $x=x_{n}\left( t\right) $ in (\ref{DDEpsi}) we get (via (\ref%
{QQx}) and (\ref{Changexz})---and (\ref{Defxnt}) implying of course $\psi
_{N}\left( x_{n}\left( t\right) ;t\right) =0$)%
\begin{eqnarray}
\dot{x}_{n} &=&\frac{(q-1)}{2q^{N}}\Bigg\{G(z_{n})\prod_{\ell =1,\ell \neq
n}^{N}K(z_{n},z_{\ell })+G\left( \frac{1}{z_{n}}\right) \prod_{\ell =1,\ell
\neq n}^{N}K\left( \frac{1}{z_{n}},\frac{1}{z_{\ell }}\right) \Bigg\}~, 
\notag  \label{xndot} \\
&\equiv &F_{n}(z_{1},\ldots ,z_{N})\equiv \hat{F}_{n}(x_{1},\ldots
,x_{N})~,~~~n=1,\ldots ,N~,
\end{eqnarray}%
where the functions $G$ and $K$ are defined by~(\ref{eq:GG}) and~(\ref{eq:KK}%
), respectively. Note that, for notational simplicity, we omitted to
indicate the $t$-dependence of $\dot{x}_{n}$, $x_{n}$, $x_{\ell }$, $z_{n}$.
Again, via $z_{s}=\exp \left( \mathbf{i}\theta _{s}\right) $ and $x_{s}=\cos
\theta _{s}$, it is clear that the right-hand side of~(\ref{xndot}) is an
even function of $\theta_s $, hence a function of $x_{s}$, where $s=1,\ldots ,N
$.

This is an interesting dynamical system, a complete investigation of which
is beyond the scope of the present paper.\ But before proceeding with our
task, let us pause and recall that the first idea to relate the zeros of
polynomials to a dynamical system goes back to Stieltjes \cite{S1885}, was
resuscitated in \cite{C1978} to identify \textquotedblleft solvable''
many-body problems (see also the extended treatment of this approach in \cite%
{C2001}), and then extensively used to obtain results concerning the zeros
of the classical polynomials and of Bessel functions, see the paper \cite%
{ABCOP1979} where several such findings are derived and reviewed. For more
recent developments along somewhat analogous lines see, for instance, \cite%
{OS2004}, \cite{vD2005} and \cite{GVZ2014}.

Here we need to focus only on the behavior of this $t$-evolution in the
infinitesimal vicinity of the equilibrium configuration $x_{n}\left(
t\right) =\bar{x}_{n}\equiv \bar{x}_{n}\left( a,b,c,d;q;N\right) $, where
the $N$ numbers $\bar{x}_{n}$ are of course the $N$ zeros of the
Askey-Wilson polynomial $p_{N}\left( a,b,c,d;q;x\right) ,$ see (\ref{xnznbar}%
). To this end we set 
\begin{subequations}
\begin{equation}
x_{n}\left( t\right) =\bar{x}_{n}+\varepsilon ~\xi _{n}\left( t\right)
~,~~~n=1,...,N~,  \label{xksiepsi}
\end{equation}%
implying of course%
\begin{equation}
\dot{x}_{n}\left( t\right) =\varepsilon ~\dot{\xi}_{n}\left( t\right)
~,~~~n=1,...,N~,  \label{xksidot}
\end{equation}%
with $\varepsilon $ infinitesimal.

It is plain that the insertion of this \textit{ansatz} in (\ref{xndot}) is
consistent to order $\varepsilon ^{0}=1$, thanks to \textbf{Proposition 2.1}.

The insertion of this \textit{ansatz} in (\ref{xndot}) yields, to order $%
\varepsilon $ (after a trivial but somewhat cumbersome computation) the
linear system of ODEs 
\end{subequations}
\begin{equation}
\dot{\xi}_{n}\left( t\right) =\sum_{m=1}^{N}\left[ M_{nm}\left( \underline{%
\bar{x}}\right) ~\xi _{m}\left( t\right) \right] ~,~~~n=1,...,N~,
\label{ksidot}
\end{equation}%
with the $N\times N$ matrix $\underline{M}\left( \underline{\bar{x}}\right)
\equiv \underline{M}\left( a,b,c,d;q;N;\underline{\bar{x}}\right) $ given by 
\begin{equation}
M_{nm}=\frac{\partial }{\partial x_{m}}\hat{F}_{n}(x_{1},\ldots ,x_{N})\Bigg|%
_{x_{s}=\bar{x}_{s}}=\frac{\partial }{\partial {z}_{m}}{F}_{n}(z_{1},\ldots
,z_{N})\frac{d{z}_{m}}{dx_{m}}\Bigg|_{z_{s}=\bar{z}_{s}},\;s=1,\ldots ,N,
\label{eq:MDefDerivFn}
\end{equation}%
which, after the substitution $\frac{dz_{m}}{dx_{m}}=\frac{2z_{m}^{2}}{%
z_{m}^{2}-1}$ (implied by $z_{m}=x_{m}+\sqrt{x_{m}^{2}-1}$) and
some trivial computations yields formula (\ref{M}). While by continuing the
expansion of the right-hand side of (\ref{xndot}) in powers of $\varepsilon $
and setting to zero the resulting coefficients of higher powers of $%
\varepsilon $ (since of course the left-hand side of (\ref{xndot}) contains
only a term of order $\varepsilon ,$ see (\ref{xksidot})), additional
formulas satisfied by the $N$ zeros $\bar{x}_{n}\equiv \bar{x}_{n}\left(
a,b,c,d;q;N\right) $ of the Askey-Wilson polynomial $P_{N}\left(
a,b,c,d;q;z\right) $ can be obtained.

The proof of \textbf{Proposition 2.2} is now a consequence of the fact that
the solution of the system of linear ODEs (\ref{ksidot}) is clearly a linear
superposition (with $t$-independent coefficients) of exponentials, $\exp
\left( \tilde{\mu}_{m}t\right) $, where the quantities $\tilde{\mu}_{m}$
(with $m=1,2,...,N$) are the $N$ eigenvalues of the $N\times N$ matrix $%
\underline{M}$; but---due to the simultaneous validity of the relations (\ref%
{psixt}), (\ref{Defxnt}) and (\ref{xksiepsi})---this solution must also be a
linear superposition (with $t$-independent coefficients) of the $t$%
-dependent quantities $c_{m}\left( t\right) .$ Hence (\ref{cmt}) implies $%
\tilde{\mu}_{m}=\mu _{m}=q^{-N}\left( 1-q^{m}\right) \left(
1-abcd~q^{2N-1-m}\right) $. \textbf{Proposition 2.2} is thereby proven.

Note that in our treatment above we implicitly assumed that the zeros $%
x_{n}\left( t\right) $ are---for all values of $t$---all different among
themselves. This is indeed the \textit{generic} situation. Any nongeneric
event---like the\ \textquotedblleft collision'' of two different zeros at
some special value of the parameter $t$---can be dealt with by appropriate
limits and in any case such possibilities---should they occur---do not
invalidate the proof of \textbf{Proposition 2.2}, as reported above.

\subsection{\textit{q}-Racah polynomials}

The following treatment is analogous---\textit{mutatis mutandis}---to that
of the previous Section 3.1, hence it shall be somewhat more terse.

Let $\Phi _{N}\left( z;t\right) \equiv \Phi _{N}\left( \alpha ,\beta ,\gamma
,\delta ;q;z;t\right) $ be a function of the variables $z$ and $t$, which
satisfies the Differential-\textit{q-}Difference Equation (D\textit{q}DE) 
\begin{subequations}
\label{PDqDEqRacah}
\begin{eqnarray}
&&\frac{\partial ~\Phi _{N}\left( z;t\right) }{\partial ~t}=\left[ \left(
q^{-N}-1\right) \left( 1-\alpha \beta q^{N+1}\right) +B\left( z\right)
+D\left( z\right) \right] ~\Phi _{N}\left( z;t\right)  \notag \\
&&-B\left( z\right) ~\Phi _{N}\left( z^{\left( +\right) };t\right) -D\left(
z\right) ~\Phi _{N}\left( z^{\left( -\right) };t\right) ~,
\end{eqnarray}%
where (see (\ref{z+-}))%
\begin{equation}
z^{\left( \pm \right) }\equiv z^{\left( \pm \right) }\left( \gamma \delta
;q;z\right) =q^{\pm 1}z\pm \left( \frac{1-q^{2}}{2q}\right) \left( z-\sqrt{%
z^{2}-4\gamma \delta q}\right)
\end{equation}%
and the functions $B\left( z\right) \equiv B\left( \alpha ,\beta ,\gamma
,\delta ;q;z\right) $ respectively $D\left( z\right) \equiv D\left( \alpha
,\beta ,\gamma ,\delta ;q;z\right) $ are defined by (\ref{B}) respectively (%
\ref{D}) with (see (\ref{Z}))%
\begin{equation}
Z\left( \gamma \delta q;z\right) =\frac{z+\sqrt{z^{2}-4\gamma \delta q}}{%
2\gamma \delta q}~.  \label{ZZ}
\end{equation}
And let us assume that this function can be expressed (as confirmed below)
by the following linear superposition with $t$-dependent coefficients $%
c_{m}\left( t\right) $ of the $(N+1)$ \textit{q}-Racah polynomials $%
R_{N-m}\left( z\right) \equiv R_{N-m}\left( \alpha ,\beta ,\gamma ,\delta
;q;z\right) $ with $m=0,1,2,...,N$: 
\end{subequations}
\begin{equation}
\Phi _{N}\left( z;t\right) =\sum_{m=0}^{N}\left[ c_{m}\left( t\right)
~R_{N-m}\left( z\right) \right] ~.  \label{Phicmt}
\end{equation}

It is then plain---see (\ref{qdiffRacah}), of course with $N$ replaced by $%
(N-m)$---that the $t$-evolution of the $(N+1)$ coefficients $c_{m}\left(
t\right) $ is characterized by the following system of ODEs, 
\begin{eqnarray}
\dot{c}_{m}\left( t\right)  &=&\left[ \left( q^{-N}-1\right) \left( 1-\alpha
\beta q^{N+1}\right) \right.   \notag \\
&&\left. -\left( q^{m-N}-1\right) ~\left( 1-\alpha \beta ~q^{N-m+1}\right) 
\right] ~c_{m}\left( t\right)   \notag \\
&=&q^{-N}~\left( 1-q^{m}\right) \left( 1-\alpha \beta ~q^{2N-m+1}\right)
~c_{m}\left( t\right) ~,  \notag \\
m &=&0,1,,2...,N~,
\end{eqnarray}%
where again the superimposed dot denotes  $t$-differentiation. Hence 
\begin{subequations}
\label{cmtqRacah}
\begin{equation}
c_{0}\left( t\right) =c_{0}\left( 0\right) ~,
\end{equation}%
\begin{equation}
c_{m}\left( t\right) =\exp \left( \lambda _{m}~t\right) ~c_{m}\left(
0\right) ~,~~~m=1,2,...,N~,  \label{cmtRacah}
\end{equation}%
where 
\begin{eqnarray}
\lambda _{m} &\equiv &\lambda _{m}\left( \alpha \beta ;q;N\right)
=q^{-N}\left( 1-q^{m}\right) \left( 1-\alpha \beta ~q^{2N-m+1}\right) ~, 
\notag \\
m &=&1,2,...,N~,
\end{eqnarray}%
implying (see (\ref{Phicmt}))%
\begin{equation}
\Phi _{N}\left( z;t\right) =c_{0}\left( 0\right) ~R_{N}\left( z\right)
+\sum_{m=1}^{N}\left[ \exp \left( \lambda _{m}~t\right) ~c_{m}\left(
0\right) ~R_{N-m}\left( z\right) \right] ~.  \label{Phizt}
\end{equation}

Next, let us introduce the $N$ zeros $z_{n}\left( t\right) \equiv
z_{n}\left( \alpha ,\beta ,\gamma ,\delta ;q;N;t\right) $ of the polynomial $%
\Phi _{N}\left( z,t\right) $ of degree $N$ in $z$, by setting 
\end{subequations}
\begin{equation}
\Phi _{N}\left( z;t\right) =C_{N}~\prod\limits_{n=1}^{N}\left[ z-z_{n}\left(
t\right) \right] ~.  \label{FacPhi}
\end{equation}%
Here $C_{N}$ is an arbitrary constant that plays no role in the following.

Let us now investigate the $t$-evolution of the $N$ zeros $z_{n}\left(
t\right) $, as implied by the Differential \textit{q}-Difference Equation (%
\ref{PDqDEqRacah}) satisfied by $\Phi _{N}\left( z;t\right) $.

It is plain from (\ref{FacPhi})---by logarithmic $t$-differentiation---that 
\begin{equation}
\frac{\partial ~\Phi _{N}\left( z;t\right) }{\partial t}=-\Phi _{N}\left(
z;t\right) ~\sum_{m=1}^{N}\frac{\dot{z}_{m}\left( t\right) }{z-z_{m}\left(
t\right) }=-C_{N}~\sum_{m=1}^{N}\left\{ \dot{z}_{m}\left( t\right)
\prod\limits_{\ell =1,~\ell \neq m}^{N}\left[ z-z_{\ell }\left( t\right) %
\right] \right\} ~.
\end{equation}%
Hence%
\begin{equation}
\left. \frac{\partial ~\Phi _{N}\left( z;t\right) }{\partial t}\right\vert
_{z=z_{n}\left( t\right) }=-C_{N}~\left\{ \dot{z}_{n}\left( t\right)
~\prod\limits_{\ell =1,~\ell \neq n}^{N}\left[ z_{n}\left( t\right) -z_{\ell
}\left( t\right) \right] \right\} ~,
\end{equation}%
and by setting $z=z_{n}\left( t\right) $ in (\ref{PDqDEqRacah}) we get (via (%
\ref{FacPhi}) implying  $\Phi _{N}\left( z_{n}\left( t\right)
;t\right) =0$) 
\begin{subequations}
\label{EvEqznRacah}
\begin{eqnarray}
&&\dot{z}_{n}=B\left( z_{n}\right) ~\left( z_{n}^{\left( +\right)
}-z_{n}\right) ~\prod\limits_{\ell =1,~\ell \neq n}^{N}\left( \frac{%
z_{n}^{\left( +\right) }-z_{\ell }}{z_{n}-z_{\ell }}\right)  \notag \\
&&+D\left( z_{n}\right) ~\left( z_{n}^{\left( -\right) }-z_{n}\right)
~\prod\limits_{\ell =1,~\ell \neq n}^{N}\left( \frac{z_{n}^{\left( -\right)
}-z_{\ell }}{z_{n}-z_{\ell }}\right) ~,  \notag \\
&&n=1,2,...,N~,
\end{eqnarray}%
where  $B\left( z\right) \equiv B\left( \alpha ,\beta ,\gamma
,\delta ;q;z\right) $ respectively $D\left( z\right) \equiv D\left( \alpha
,\beta ,\gamma ,\delta ;q;z\right) $ are defined by (\ref{B}) respectively (%
\ref{D}) with (\ref{Z}), and%
\begin{equation}
z_{n}^{\left( \pm \right) }=q^{\pm 1}z_{n}\pm \left( \frac{1-q^{2}}{2q}%
\right) \left( z_{n}-\sqrt{z_{n}^{2}-4\gamma \delta q}\right)
~,~~~n=1,...,N~.
\end{equation}%
Note that we  omitted  the arguments of $z_{n}\equiv
z_{n}\left( \alpha ,\beta ,\gamma ,\delta ;q;N;t\right) $ and, likewise, $%
z_{n}^{\left( \pm \right) }\equiv z_{n}^{\left( \pm \right) }\left( \alpha
,\beta ,\gamma ,\delta ;q;N;t\right) $.

Let us again mention that this set of nonlinear ODEs, (\ref{EvEqznRacah}),
is an interesting dynamical system, a complete investigation of which is
however beyond the scope of the present paper. Here we need to focus only on
its behavior in the infinitesimal vicinity of its equilibrium configuration $%
z_{n}\left( t\right) =\bar{z}_{n}\equiv \bar{z}_{n}\left( \alpha ,\beta
,\gamma ,\delta ;q;N\right) $, where the $N$ numbers $\bar{z}_{n}$ are now
of course the $N$ zeros of the \textit{q}-Racah polynomial $R_{N}\left(
\alpha ,\beta ,\gamma ,\delta ;q;z\right) $. To this end we set 
\end{subequations}
\begin{subequations}
\begin{equation}
z_{n}\left( t\right) =\bar{z}_{n}+\varepsilon ~\zeta _{n}\left( t\right)
~,~~~n=1,...,N~,  \label{zitaepsi}
\end{equation}%
implying of course%
\begin{equation}
\dot{z}_{n}\left( t\right) =\varepsilon ~\dot{\zeta}_{n}\left( t\right)
~,~~~n=1,...,N~,
\end{equation}%
with $\varepsilon $ infinitesimal.

It is plain that the insertion of this \textit{ansatz} in (\ref{EvEqznRacah}%
) is consistent to order $\varepsilon ^{0}=1$, thanks to \textbf{Proposition
2.3}.

The insertion of this \textit{ansatz} in (\ref{EvEqznRacah}) yields, to
order $\varepsilon $ (after a trivial but somewhat cumbersome computation)
the \textit{linear} system of ODEs 
\end{subequations}
\begin{equation}
\dot{\zeta}_{n}\left( t\right) =\sum_{m=1}^{N}\left[ L_{nm}\left( \underline{%
\bar{z}}\right) ~\zeta _{m}\left( t\right) \right] ~,~~~n=1,...,N~,
\label{zittandot}
\end{equation}%
with the $N\times N$ matrix $\underline{L}\left( \underline{\bar{z}}\right)
\equiv \underline{L}\left( \alpha ,\beta ,\gamma ,\delta ;q;N;\underline{%
\bar{z}}\right) $ defined by (\ref{L}). While by continuing the expansion of
the right-hand side of (\ref{EvEqznRacah}) in powers of $\varepsilon $ and
setting to zero the resulting coefficients of higher powers of $\varepsilon $%
, additional formulas satisfied by the $N$ zeros $\bar{z}_{n}\equiv \bar{z}%
_{n}\left( \alpha ,\beta ,\gamma ,\delta ;q;N\right) $ of the \textit{q}%
-Racah polynomial $R_{N}\left( \alpha ,\beta ,\gamma ,\delta ;q;z\right) $
can be obtained.

The proof of \textbf{Proposition 2.4} is now a consequence of the fact that
the solution of the system of \textit{linear} ODEs (\ref{zittandot}) is
clearly a \textit{linear} superposition (with $t$-independent coefficients)
of exponentials, $\exp \left( \tilde{\lambda}_{m}t\right) $, where the
quantities $\tilde{\lambda}_{m}$ (with $m=1,2,...,N$) are the $N$
eigenvalues of the $N\times N$ matrix $\underline{L}$; but---due to the
simultaneous validity of the relations (\ref{Phizt}), (\ref{FacPhi}) and (%
\ref{zitaepsi})---this solution must also be a \textit{linear} superposition
(with $t$-independent coefficients) of the $t$-dependent quantities $%
c_{m}\left( t\right) .$ Hence (\ref{cmtqRacah}) implies $\tilde{\lambda}%
_{m}=\lambda _{m}=q^{-N}\left( 1-q^{m}\right) \left( 1-\alpha \beta
~q^{2N-m+1}\right) ,$ $m=1,2,...,N$. \textbf{Proposition 2.4} is thereby
proven.

\section{Appendix: Standard definitions and properties of the Askey-Wilson
and \textit{q}-Racah polynomials}

In this Appendix we report for the convenience of the reader, and also to
identify the notation used throughout this paper, a number of standard
formulas associated with the Askey-Wilson and \textit{q}-Racah polynomials.
We generally report these formulas from the standard compilation \cite{KS},
the formulas of which are identified by the notations (KS-X), where X stands
here for the notation appropriate to identify equations in this compilation;
in some cases we add certain immediate consequences of these formulas which
are not explicitly displayed in this compilation nor (to the best of our
knowledge) elsewhere.

The \textit{q}-Pochhammer symbol is defined as follows: 
\begin{subequations}
\begin{equation}
\left( c;q\right) _{0}=1~;~~~\left( c;q\right) _{n}=\left( 1-c\right)
~\left( 1-cq\right) \cdot \cdot \cdot \left( 1-cq^{n-1}\right) ~~~\text{%
for~~~}n=1,2,3,...~,  \label{q-Poch}
\end{equation}%
and we also use occasionally the synthetic notation%
\begin{equation}
\left( c_{1};q\right) _{k}\left( c_{2};q\right) _{k}\cdot \cdot \cdot \left(
c_{r};q\right) _{k}\equiv \left( c_{1},c_{2},...,c_{r};q\right) _{k}~,~\ ~%
\text{for~~~}r=0,1,2,...~.  \label{Genq-Poch}
\end{equation}

The generalized basic hypergeometric function is defined as follows (see
(KS-0.4.2)): 
\end{subequations}
\begin{eqnarray}
&&_{r+1}\phi _{s}\left( \left. 
\begin{array}{c}
a_{0},a_{1},...,a_{r} \\ 
b_{1},b_{2},...,b_{s}%
\end{array}%
\right\vert q;z\right)  \notag \\
&=&\sum_{k=0}^{\infty }\left[ \frac{\left( a_{0},a_{1},...,a_{r};q\right)
_{k}}{\left( b_{1},b_{2},...,b_{s};q\right) _{k}}\left( -1\right) ^{\left(
s-r\right) k}~q^{\left( s-r\right) k\left( k-1\right) /2}~\frac{z^{k}}{%
\left( q;q\right) _{k}}\right] ~.  \label{BasicHypFunct}
\end{eqnarray}%
Here $r$ and $s$ are two arbitrary \textit{nonnegative} integers, but in the
following consideration shall be restricted to $r=s=3$.

The basic hypergeometric function (\ref{BasicHypFunct}) becomes a \textit{%
polynomial} in $z$ of degree $N$ if one of the parameters $a_{n}$ has the
values $q^{-N}$ with $N$ a positive integer---say, $a_{0}=$ $q^{-N}$
(without loss of generality, since its definition (\ref{BasicHypFunct})
implies that the basic generalized hypergeometric function is invariant
under permutations of the $r+1$ parameters $a_{j}$ as well as under
permutations of the $s$ parameters $b_{k}$)---provided no one of the other
parameters $a_{j}$ equals $q^{-\nu }$ with $\nu $ a negative integer smaller
in modulus than $N$, and no one of the parameters $b_{k}$ equals $q^{-\nu }$
with $\nu $ a negative integer (as we hereafter assume). This is of course a
simple consequence of the fact that, if $N$ is a positive integer, $\left(
q^{-N};q\right) _{n}$ vanishes for $n>N,$ see (\ref{q-Poch}). Note that in
this case the basic hypergeometric function is also a polynomial in each of
the $r$ parameters $a_{j}$ with $j=1,2,...,r.$

\subsection{Formulas for the Askey-Wilson polynomials}

Hereafter $\mathbf{i}$ denotes the imaginary unit, $\mathbf{i}^{2}=-1$.

The Askey-Wilson polynomial $p_{N}(a,b,c,d;q;x)$ with $x=\cos  \theta$ 
is defined as follows (see (KS-3.1.1), and note some minor
notational changes): 
\begin{subequations}
\label{AWpolx}
\begin{eqnarray}
&&p_{N}(a,b,c,d;q;\cos \theta  )=\frac{\left(
ab,ac,ad;q\right) _{N}}{a^{N}}\cdot   \notag \\
&&\cdot _{4}\phi _{3}\left( \left. 
\begin{array}{c}
q^{-N},~abcd~q^{N-1},~a~\exp \left( \mathbf{i}\theta \right) ,~a~\exp \left(
-\mathbf{i}\theta \right)  \\ 
ab,~ac,~ad%
\end{array}%
\right\vert q;q\right) ~,
\end{eqnarray}%
or, equivalently but more explicitly,%
\begin{eqnarray}
&&p_{N}(a,b,c,d;q;x)=\frac{\left( ab,ac,ad;q\right) _{N}~}{a^{N}}\cdot  
\notag \\
&&\cdot \sum_{m=0}^{N}\left[ \frac{q^{m}\left( q^{-N};q\right) _{m}~\left(
abcd~q^{N-1};q\right) _{m}}{\left( q;q\right) _{m}~\left( ab;q\right)
_{m}~\left( ac;q\right) _{m}~\left( ad;q\right) _{m}}~\left\{ a;q;x\right\}
_{m}\right] ~,  \notag \\
&&  \label{AWPol}
\end{eqnarray}%
where we introduced the new (modified \textit{q}-Pochhammer) symbol defined
as follows: 
\end{subequations}
\begin{eqnarray}
\left\{ a;q;x\right\} _{0} &=&1~;  \notag \\
\left\{ a;q;x\right\} _{m} &=&\left( 1+a^{2}-2ax\right) ~\left(
1+q^{2}a^{2}-2aqx\right) \cdot \cdot \cdot \left(
1+a^{2}q^{2(m-1)}-2aq^{m-1}x\right) ~,  \notag \\
m &=&1,2,3,...~.  \label{ModqPoch}
\end{eqnarray}%
It is plain from this formula that $\left\{ a;q;x\right\} _{m}$ is a
polynomial of degree $m$ in $x$ (and also of degree $2m$ in $a$), hence that
the Askey-Wilson polynomials $p_{N}(a,b,c,d;q;x)$\textit{\ }are indeed
polynomials of degree $N$ in $x$ (see (\ref{AWPol})).

\textit{Notational remark}. For notational simplicity we often omit to
indicate explicitly the dependence on the $5$ parameters $a$, $b,$ $c,$ $d,$ 
$q$---or on some of them---provided this entails no ambiguity. $\square $

Let us also recall the\ related \textit{rational} function of $z$ defined as
follows: 
\begin{eqnarray}
&&P_{N}\left( a,b,c,d;q;z\right) =\frac{\left( ab,ac,ad;q\right) _{N}}{a^{N}}%
\cdot  \notag \\
&&\cdot _{4}\phi _{3}\left( \left. 
\begin{array}{c}
q^{-N},~abcd~q^{N-1},~az,~a/z \\ 
ab,~ac,~ad%
\end{array}%
\right\vert q;q\right) ~,  \label{PNz}
\end{eqnarray}%
see the formula in \cite{KS} preceding (KS-3.1.7). It is plain that this
amounts to the change of variables%
\begin{equation}
x=\cos \theta =\frac{z^{2}+1}{2~z}~,~~\ z=e^{i\theta }=x+\sqrt{x^{2}-1}~,
\label{Changexz}
\end{equation}%
corresponding to the relations 
\begin{subequations}
\label{PpAW}
\begin{equation}
P_{N}\left( a,b,c,d;q;z\right) =p_{N}\left( a,b,c,d;q;\frac{z^{2}+1}{2~z}%
\right) ~,
\end{equation}%
\begin{equation}
p_{N}\left( a,b,c,d;q;x\right) =P_{N}\left( a,b,c,d;q;x+\sqrt{x^{2}-1}%
\right) ~.
\end{equation}%
Note that although the square root $\sqrt{x^{2}-1}$ has two possible values,
the definition of the function $P_{N}\left( z\right) $, see (\ref{PNz}) and (%
\ref{q-Poch}), implies that the choice of the square root is irrelevant
because $\left( x+\sqrt{x^{2}-1}\right) \left( x-\sqrt{x^{2}-1}\right) =1$.
Hence if we denote as $\bar{x}_{n}$ the $N$ zeros of the Askey-Wilson
polynomial $p_{N}\left( a,b,c,d;q;x\right) ,$%
\end{subequations}
\begin{subequations}
\label{xnznbar}
\begin{equation}
p_{N}\left( a,b,c,d;q;\bar{x}_{n}\right) =0~,~\ ~n=1,2,...,N~,
\label{pNzeros}
\end{equation}%
it is plain that the rational function $P_{N}(a,b,c,d;q;z)$ features the $2N$
zeros $\bar{z}_{n}=\bar{x}_{n}+\sqrt{\bar{x}_{n}^{2}-1},$ 
\begin{equation}
P_{N}\left( a,b,c,d;q;\bar{z}_{n}\right) =0~,~\ ~n=1,2,...,N~.
\label{PNzeros}
\end{equation}%
In the following we will use whichever one of the two assignments of each square root $%
\sqrt{\bar{x}_{n}^{2}-1}$, $n=1,\ldots ,N$, is more convenient in order to
simplify the following formulas; of course the zeros $\bar{z}_{n}$ are
functions of the $6$ parameters $a$, $b$, $c$, $d$, $q$, $N$, 
\begin{equation}
\bar{x}_{n}\equiv \bar{x}_{n}\left( a,b,c,d;q;N\right) ~,~~~\bar{z}%
_{n}\equiv \bar{z}_{n}\left( a,b,c,d;q;N\right) ~,~~~n=1,2,...,N~,
\end{equation}%
and they are related by the formulas 
\begin{equation}
\bar{x}_{n}=\cos \bar{\theta}_{n}=\frac{\bar{z}_{n}^{2}+1}{2~\bar{z}_{n}}%
~,~~\ \bar{z}_{n}=e^{i\bar{\theta}_{n}}=\bar{x}_{n}+\sqrt{\bar{x}_{n}^{2}-1}%
~,~~~n=1,2,...,N~.  \label{xnzn}
\end{equation}%
Note that occasionally we abuse language by referring both to the $N$
(generally \textit{complex}) numbers $\bar{x}_{n}$ and to the $2N$
(generally \textit{complex}) numbers $\bar{z}_{n}$ as \textit{zeros} of the
Askey-Wilson polynomial (of degree $N$), although of course only the $N$
numbers $\bar{x}_{n}$ are indeed $N$ zeros of the Askey-Wilson \textit{%
polynomial} $p_{N}\left( a,b,c,d;q;x\right) $, see (\ref{pNzeros}), while
the $2N$ numbers $\bar{z}_{n}$ defined by  (\ref{xnzn}), in
terms of the $N$ numbers $\bar{x}_{n}$, are the zeros of the 
\textit{rational} function $P_{N}\left( a,b,c,d;q;z\right) $, see (\ref%
{PNzeros}).

The rational function $P_{N}\left( a,b,c,d;q;z\right) $ satisfies the
following $q$-difference equation (see (KS-3.1.7)): 
\end{subequations}
\begin{subequations}
\label{qDE}
\begin{equation}
Q~P_{N}\left( a,b,c,d;q;z\right) =\left( q^{-N}-1\right) ~\left(
1-abcd~q^{N-1}\right) ~P_{N}\left( a,b,c,d;q;z\right) ~,  \label{Qeigen}
\end{equation}%
where the \textit{q}-difference operator $Q$ is defined as follows:%
\begin{equation}
Q~f\left( z\right) =\left[ A\left( z\right) ~\Delta _{q}^{\left( +\right)
}+A\left( z^{-1}\right) ~\Delta _{q}^{\left( -\right) }-A\left( z\right)
-A\left( z^{-1}\right) \right] ~f\left( z\right) ~.  \label{Q}
\end{equation}%
Here $f\left( z\right) $ is an arbitrary function of the variable $z,$ the
operators $\Delta _{q}^{\left( \pm \right) }$ act as follows on $f\left(
z\right) ,$%
\begin{equation}
\Delta _{q}^{\left( \pm \right) }~f\left( z\right) =f\left( q^{\pm
1}~z\right) ~,  \label{deltaplusminus}
\end{equation}%
and the function $A\left( z\right) $ is defined (here and throughout) as
follows:%
\begin{equation}
A\left( z\right) \equiv A\left( a,b,c,d;q;z\right) =\frac{\left( 1-az\right)
~\left( 1-bz\right) ~\left( 1-cz\right) ~\left( 1-dz\right) }{\left(
1-z^{2}\right) ~\left( 1-qz^{2}\right) }~.  \label{A}
\end{equation}%
These formulas indicate that the operator $Q$---the definition of which
features (symmetrically) the $4$ parameters $a$, $b$, $c$, $d$ and moreover
the parameter $q$ (but not the parameter $N$)---has the rational functions $%
P_{N}(a,b,c,d;q;z)$ (for all positive integer values of $N$) as its
eigenfunctions, with corresponding eigenvalues $\left( q^{-N}-1\right)
~\left( 1-abcd~q^{N-1}\right) ,$ see (\ref{Qeigen}).

Note that, by setting $z=\bar{z}_{n}$ in (\ref{Qeigen}) one obtains the
following set of algebraic equations satisfied by the zeros of the
Askey-Wilson polynomial of degree $N$: 
\end{subequations}
\begin{eqnarray}
&&A\left( \bar{z}_{n}\right) ~P_{N}(a,b,c,d;q;q~\bar{z}_{n}) +A\left( \frac{1%
}{\bar{z}_{n}}\right) ~P_{N}(a,b,c,d;q;\frac{\bar{z}_{n}}{q})=0~,  \notag \\
n &=&1,...,N~.  \label{Prop21}
\end{eqnarray}%
This observation is instrumental to prove \textbf{Proposition 2.2} (see
Section 3); it corresponds, via (\ref{A}) and (\ref{xnzn}), to \textbf{%
Proposition 2.1}---which has been displayed in Section 2 because we did not
find any previous mention of this rather trivial finding in the literature.
Although in fact---after this paper of ours was posted in the web as
arXiv:1410.0549---Jan Felipe van Diejen kindly brought to our attention that
essentially the same result was already reported in his paper \cite{vD2005},
see Theorem 2 and eq. (3.1b) there; and likewise our finding concerning the
zeros of the Wilson polynomials (see eq. (40) in our previous paper \cite%
{BC2014a}) also appears there, see Theorem 3 in \cite{vD2005}.

\subsection{Formulas for the \textit{q}-Racah polynomials}

The \textit{q}-Racah polynomial $R_{N}(\alpha ,\beta ,\gamma ,\delta;q;z)$
is a polynomial of degree $N$ in $z$, 
\begin{subequations}
\label{qRacah}
\begin{equation}
z\equiv z\left( \gamma \delta ;q;x\right) =q^{-x}+\gamma \delta q^{x+1}~,
\label{zqRacah}
\end{equation}%
defined as follows in terms of the generalized basic hypergeometric function
(see (KS-3.2.1), and note some minor notational changes): 
\begin{equation}
R_{N}(\alpha ,\beta ,\gamma ,\delta ;q;z)=_{4}\phi _{3}\left( \left. 
\begin{array}{c}
q^{-N},~\alpha \beta q^{N+1},~q^{-x},~\gamma \delta q^{x+1} \\ 
\alpha q,~\beta \delta q,~\gamma q%
\end{array}%
\right\vert q;q\right) ~,  \label{qRacaha}
\end{equation}%
or, equivalently but more explicitly,%
\begin{eqnarray}
&&R_{N}(\alpha ,\beta ,\gamma ,\delta ;q;z)=  \notag \\
&=&\sum_{m=0}^{N}\left[ \frac{q^{m}\left( q^{-N};q\right) _{m}\left( \alpha
\beta ~q^{N+1};q\right) _{m}\left( q^{-x};q\right) _{m}\left( \gamma \delta
q^{x+1};q\right) _{m}}{\left( q;q\right) _{m}\left( \alpha q;q\right)
_{m}\left( \beta \delta q;q\right) _{m}\left( \gamma q;q\right) _{m}}\right]
~.  \label{qRacah2}
\end{eqnarray}%
The fact that $R_{N}(z)\equiv R_{N}(\alpha ,\beta ,\gamma ,\delta ;q;z)$ is
indeed a polynomial of degree $N$ in $z$ is implied by the last formula and
by the identity (valid via (\ref{zqRacah})), 
\end{subequations}
\begin{equation}
\left( q^{-x};q\right) _{m}\left( \gamma \delta q^{x+1};q\right)
_{m}=\prod\limits_{s=0}^{m-1}\left( 1-zq^{s}+\gamma \delta q^{2s+1}\right) ~.
\end{equation}

\textbf{Remark A.1}. Hereafter the $5$ parameters $\alpha $, $\beta $, $%
\gamma $, $\delta $, $q\neq 1$ are arbitrary numbers (possibly \textit{%
complex}); note that this entails a somewhat more general definition of 
\textit{q}-Racah polynomials than the standard one, see (KS-3.2.1), because
it does \textit{not} require the Diophantine restriction on one of the $3$
parameters, $\alpha q$ or $\beta \delta q$ or $\gamma q$, see the second
(unnumbered) equation after (KS-3.2.1) in \cite{KS}. Indeed, while this
restriction is required for the validity of many of the properties of 
\textit{q}-Racah polynomials reported in \cite{KS}, it is not required for
the \textit{q}-difference equation satisfied by these polynomials, see
(KS-3.2.6) and immediately below, which is the only property of these
polynomials that we use in order to prove the properties of the zeros of
these polynomials reported in this paper. $\square $ 

The \textit{q}-Racah polynomial $R_{N}(z)\equiv R_{N}(\alpha ,\beta ,\gamma
,\delta ;q;z)$ satisfies the following \textit{q}-difference equation: 
\begin{subequations}
\label{qdiffRacah}
\begin{eqnarray}
&&B\left( z\right) ~R_{N}(z^{\left( +\right) })-\left[ B\left( z\right)
+D\left( z\right) \right] ~R_{N}(z)+D\left( z\right) ~R_{N}(z^{\left(
-\right) })  \notag \\
&=&\left( q^{-N}-1\right) \left( 1-\alpha \beta q^{N+1}\right) ~R_{N}(z)~,
\label{qdiffRacaha}
\end{eqnarray}%
where%
\begin{equation}
z^{\left( \pm \right) }=z\left( x\pm 1\right) =q^{\pm 1}z\pm \left( \frac{%
1-q^{2}}{2q}\right) \left[ z-\sqrt{z^{2}-4\gamma \delta q}\right]
\label{z+-}
\end{equation}%
and%
\begin{equation}
B\left( z\right) =\frac{\left[ 1-\alpha qZ\left( q;z\right) \right] ~\left[
1-\beta \delta qZ\left( q;z\right) \right] ~\left[ 1-\gamma qZ\left(
q;z\right) \right] ~\left[ 1-\gamma \delta qZ\left( q;z\right) \right] }{%
\left[ 1-\gamma \delta qZ^{2}\left( q;z\right) \right] ~\left[ 1-\gamma
\delta q^{2}Z^{2}\left( q;z\right) \right] }~,  \label{B}
\end{equation}%
\begin{equation}
D\left( z\right) =\frac{q~\left[ 1-Z\left( \gamma \delta q;z\right) \right] ~%
\left[ 1-\delta Z\left( \gamma \delta q;z\right) \right] ~\left[ \beta
-\gamma Z\left( \gamma \delta q;z\right) \right] ~\left[ \alpha -\gamma
\delta Z\left( \gamma \delta q;z\right) \right] }{\left[ 1-\gamma \delta
Z^{2}\left( \gamma \delta q;z\right) \right] ~\left[ 1-\gamma \delta
qZ^{2}\left( \gamma \delta q;z\right) \right] }~,  \label{D}
\end{equation}%
where%
\begin{equation}
Z\left( \gamma \delta q;z\right) =q^{x}=\frac{z+\sqrt{z^{2}-4\gamma \delta q}%
}{2\gamma \delta q}~.  \label{Z}
\end{equation}%
In (\ref{z+-}) and (\ref{Z}) the determination of the square root is
irrelevant; of course provided the \textit{same} determination is used
everywhere.

It is easily seen that these formulas correspond to (KS-3.2.6) via (\ref%
{zqRacah}).

It is plain from this formula that, if $\bar{z}_{n}$ are the $N$ zeros of
the \textit{q}-Racah polynomial of order $N$, 
\end{subequations}
\begin{subequations}
\begin{equation}
R_{N}(\bar{z}_{n})=0~,~~~n=1,...,N~,
\end{equation}%
\begin{equation}
R_{N}(z)=C_{N}~\prod\limits_{n=1}^{N}\left( z-\bar{z}_{n}\right) ~,
\label{FacRacah}
\end{equation}%
where $C_N$ is a constant. Formula (\ref{qdiffRacaha}) implies the relation 
\end{subequations}
\begin{subequations}
\begin{equation}
B\left( \bar{z}_{n}\right) ~R_{N}(\bar{z}_{n}^{\left( +\right) })+D\left( 
\bar{z}_{n}\right) ~R_{N}(\bar{z}_{n}^{\left( -\right) })=0~, 
\label{48a}
\end{equation}%
where of course (see (\ref{z+-}))%
\begin{equation}
\bar{z}_{n}^{\left( \pm \right) }=q^{\pm 1}\bar{z}_{n}\pm \left( \frac{%
1-q^{2}}{2q}\right) \left( \bar{z}_{n}-\sqrt{\bar{z}_{n}^{2}-4\gamma \delta q%
}\right) ~;\ 
\end{equation}%
and this formula coincides with (\ref{Prop23a}) and, via (\ref{FacRacah}),
with (\ref{Prop23b}). \textbf{Proposition 2.3} is thereby proven.

\section{Acknowledgements}

One of us (OB) would like to acknowledge with thanks the hospitality of the
Physics Department of the University of Rome ``La Sapienza'' on the occasion
of three two-week visits there in June 2012, May 2013 and June-July 2014;
the results reported in this paper were obtained during the last of these
visits. The other one (FC) would like to acknowledge with thanks the
hospitality of Concordia College for a one-week visit there in November 2013.

\end{subequations}

\end{document}